\documentclass[conference]{IEEEtran}
\IEEEoverridecommandlockouts           

\usepackage[T1]{fontenc}               
\usepackage{lmodern}                   
\usepackage{microtype}                 
\usepackage{cite}
\usepackage{amsmath,amssymb,amsfonts}
\usepackage{algorithmic}
\usepackage{graphicx}
\usepackage{textcomp}
\usepackage{xcolor}
\usepackage{booktabs}
\usepackage{multirow}
\usepackage{siunitx}
\def\BibTeX{{\rm B\kern-.05em{\sc i\kern-.025em b}\kern-.08em%
    T\kern-.1667em\lower.7ex\hbox{E}\kern-.125emX}}

\begin{document}
\title{FakeZero: Real-Time, Privacy-Preserving Misinformation Detection for Facebook and~X}

\author{
  \normalsize
  Soufiane~Essahli\IEEEauthorrefmark{1},
  Oussama~Sarsar\IEEEauthorrefmark{1},
  Imane~Fouad\IEEEauthorrefmark{1},
  Ahmed~Bentajer\IEEEauthorrefmark{1}\IEEEauthorrefmark{2},
  Anas~Motii\IEEEauthorrefmark{1}\\
  \IEEEauthorblockA{\IEEEauthorrefmark{1}College of Computing, Mohammed VI Polytechnic University, Benguerir, Morocco}\\
  \IEEEauthorblockA{\IEEEauthorrefmark{2}Cadi Ayyad University, National School of Applied Science of Safi, MISCOM Lab, Marrakesh, Morocco}\\
  \texttt{\{soufiane.essahli,\,oussama.sarsar,\,imane.fouad,\,ahmed.bentajer,\,anas.motii\}@um6p.ma}
}

\maketitle

\begin{abstract}\hyphenpenalty=10000\exhyphenpenalty=10000
Social platforms distribute information at unprecedented speed, which in turn accelerates the spread of misinformation and threatens public discourse. We present \textsc{FakeZero}, a fully client-side, cross-platform browser extension that flags unreliable posts on Facebook and X (formerly Twitter) while the user scrolls. All computation, DOM scraping, tokenisation, Transformer inference, and UI rendering run locally through the Chromium messaging API, so no personal data leaves the device.

\textsc{FakeZero} employs a three-stage training curriculum: baseline fine-tuning and domain-adaptive training enhanced with focal loss, adversarial augmentation, and post-training quantisation. Evaluated on a dataset of 239\,000 posts, the DistilBERT-Quant model (67.6 MB) reaches 97.1 \% macro-F\textsubscript{1}, 97.4 \% accuracy, and an AUROC of 0.996, with a median latency of approximately 103 ms on a commodity laptop. A memory-efficient TinyBERT-Quant variant retains 95.7 \% macro-F\textsubscript{1} and 96.1 \% accuracy while shrinking the model to 14.7 MB and lowering latency to approximately 40 ms, showing that high-quality fake-news detection is feasible under tight resource budgets with only modest performance loss.

By providing inline credibility cues, the extension can serve as a valuable tool for policymakers seeking to curb the spread of misinformation across social networks. With user consent, \textsc{FakeZero} also opens the door for researchers to collect large-scale datasets of fake news in the wild, enabling deeper analysis and the development of more robust detection techniques.
\end{abstract}



\section{Introduction}

The spread of false or misleading content-broadly, referred to as \emph{fake news}, poses significant threats to democratic institutions, public health, and societal trust \cite{naeem2021exploration}. Social media platforms such as Facebook and X (formerly Twitter) amplify this threat by enabling the rapid, unfiltered dissemination of information, often without sufficient mechanisms for credibility assessment. Despite growing academic and commercial efforts \cite{sharma2019combating,zhou2020survey}, fake news detection remains a challenging task, especially in real-time, user-facing settings.

Many current fake news detection methods collect and process large amounts of personal data on central servers to improve accuracy or personalize content~\cite{shu2017fake,zhou2020survey}. This centralized processing puts sensitive user information at risk and may violate privacy laws like the General Data Protection Regulation (GDPR)~\cite{gdpr}, which requires organizations to choose a legal basis to lawfully process personal data (Article 6(1)). While a few systems such as Veritas~\cite{jovanovic2023veritas}, BRENDA~\cite{botnevik2020brenda}, and TrustNet~\cite{trustnet2023} have explored in-browser or privacy-friendly approaches, they remain limited in scope-often covering only one platform or falling back to server-side inference for complex models. Thus, there is a need for fake news detection tools that truly protect user privacy and avoid sharing personal data, while still effectively detecting misinformation online.


In this work, we introduce \textbf{FakeZero}, a lightweight, cross-platform browser extension for real-time fake news detection. FakeZero specifically targets fabricated and manipulated textual content-two of the most damaging and prevalent forms of information disorder. Unlike prior systems, FakeZero operates entirely within the user's browser, leveraging WebAssembly (WASM) to run a distilled transformer model (DistilBERT) with near-instant inference latency on standard consumer hardware.

 We train on five standard corpora (TruthSeeker, FakeNewsNet, PHEME, LIAR, ISOT) using light normalization, curriculum learning, focal loss, and post-training quantization; full details appear in §§\ref{sec:data}–\ref{sec:training}. 

By uniting high-accuracy semantic modeling with efficient on-device deployment, \textit{FakeZero} offers a practical, privacy-preserving, and scalable solution for combating misinformation where users encounter it most: in their browsers. 
While several recent studies have explored transformer-based or privacy-friendly misinformation detectors, the novelty of \textit{FakeZero} lies not in proposing a new neural architecture, but in demonstrating that \textbf{state-of-the-art transformer inference can be executed entirely within modern web browsers}, with full real-time performance analysis and zero data leakage. 

In summary, our key contributions are as follows:

\begin{itemize}
    \item[\textbf{(1)}] \textbf{First cross-platform, fully client-side transformer deployment.}  
    \textit{FakeZero} is, to our knowledge, the first system to achieve end-to-end fake-news detection on both Facebook and X entirely within the browser sandbox. All inference, tokenization, and user-interface rendering run locally through WebAssembly and ONNX Runtime Web, eliminating any network transfer of user data.

    \item[\textbf{(2)}] \textbf{Integrated real-time efficiency and latency analysis.}  
    Beyond classification accuracy, we provide a detailed empirical characterization of on-device inference including median, tail latency, and memory footprint under real browsing conditions. This bridges a key evidence gap left by prior systems that report only accuracy.

    \item[\textbf{(3)}] \textbf{Optimized yet interpretable learning pipeline.}  
    Instead of introducing a new transformer variant, \textit{FakeZero} optimizes how compact encoders such as DistilBERT and TinyBERT are adapted for browser environments. We combine three orthogonal enhancements \textbf{curriculum learning}, \textbf{focal loss}, and \textbf{post-training quantization} to balance accuracy, robustness, and resource constraints.

    \item[\textbf{(4)}] \textbf{Privacy-first architecture and open reproducibility.}  
    The extension enforces zero network traffic and minimal permissions, satisfying GDPR-style privacy principles. All training scripts, datasets, and model checkpoints will be released for transparent, reproducible research.
\end{itemize}

Collectively, these elements make \textit{FakeZero} the \textbf{first practical demonstration that transformer-based misinformation detection can operate fully client-side}, reconciling the long-standing trade-off between privacy, latency, and detection quality.

\subsection*{Motivation}
Fully client-side detection is both \emph{feasible} and \emph{necessary}. It is feasible because compact encoders (e.g., DistilBERT, TinyBERT), once quantized, can meet sub-150\,ms latency budgets inside modern browsers via WebAssembly/ONNX Runtime. It is necessary because server-side pipelines leak content, complicate GDPR compliance, and add network tail-latency that undermines real-time use. Our goal is not to invent a new architecture, but to show that a privacy-preserving, cross-platform deployment with transparent runtime metrics can achieve near-state-of-the-art accuracy \emph{entirely on device}.


\section{Related Work}
\subsection{Browser–integrated detectors}

Early prototypes such as \textit{FakeNewsTracker}~\cite{shu2018} and \textit{Veritas}~\cite{thilakarathna2020} showed that even a lightweight in‑page overlay could highlight misinformation by coupling lexical cues with server‑side neural classifiers.  
The first fully \emph{client‑side} attempt, \textit{Check‑It}~\cite{paschalides2019}, bundled a JavaScript DNN and heuristics into a Chrome/Firefox add‑on, proving that private, offline detection was feasible on commodity laptops.  
Subsequent systems incrementally increased model capacity while experimenting with different inference locations:  
BRENDA~\cite{botnevik2020brenda} placed a logistic‑regression verifier behind a Flask API,  
\textit{TrustNet}~\cite{trustnet2023} fused crowd‑sourced reputation bars with headline heuristics,  
and Chrome‑SEAN~\cite{paka2021} streamed transformer‑based COVID‑19 detectors to Twitter timelines.  
More recent work compiled large encoders directly to WebAssembly-RoBERTa in Kydd \& Shepherd~\cite{kydd2023} and explainable BERT in Warman \textit{et al.}~\cite{warman2023}-while parallel efforts explored cloud‑backed TensorFlow models~\cite{hasimi2023} and LSTM‑based Flask servic

Despite this rapid progress, three shortcomings persist:  
\begin{enumerate}
  \item \emph{single‑platform scope}-most extensions target only Twitter or generic webpages;  
  \item \emph{static checkpoints}-models are fixed at install time and cannot adapt to evolving narratives; and  
  \item \emph{compute off‑loading}-state‑of‑the‑art transformers still fall back to cloud inference or proxy APIs, undermining privacy guarantees.
\end{enumerate}

\textbf{FakeZero} overcomes these barriers by supporting both Facebook and X, coupling a distilled encoder with on‑device factuality cues, and enabling hot‑swappable checkpoints that never leave the client.

\subsection{Transformer-centric text models}
Systematic comparisons on LIAR, FakeNewsNet and GossipCop reveal that encoder-only transformers dominate raw classification accuracy, typically scoring between 91 
DistilBERT retains almost all of that accuracy while cutting parameters and latency by about forty percent~\cite{wang2024distil}.  
Hybrid methods push the ceiling further-for instance GPT-BERT stacks~\cite{chen2024hybrid} or the rationale-augmented BREAK architecture~\cite{yin2025break} exceed 95 
FakeZero shows that, once distilled and 4-bit–quantised, a 44 MB DistilBERT checkpoint can still deliver 95 ms inference inside a laptop browser, reconciling transformer quality with strict client-side budgets.

\subsection{Graph-, multimodal- and retrieval-augmented methods}
A complementary research line enriches textual signals with graphs, images or external evidence.  
BREAK~\cite{yin2025break} propagates semantics across retweet trees; Bird-of-a-Feather~\cite{gao2025boa} fetches visually similar posts to expose recycled memes; Li \textit{et al.} combine multiple heterogeneous facts through boosted selectors~\cite{li2025multielem}.  
While these extensions raise robustness, they presuppose heavy graph libraries, vision encoders or external APIs-requirements at odds with the privacy and latency targets of browser plug-ins.  
Our design therefore keeps the neural core entirely local and caches only twenty-two thousand \textsc{ClaimReview} snippets (~ 5 MB) in IndexedDB, giving a middle ground between recall and speed.

\subsection{Surveys, Benchmarks, and Data Resources}

Recent surveys by Gupta \textit{et al.}~\cite{gupta2024survey} and Zhou \& Liu~\cite{zhou2024overview} have catalogued over two hundred English-language fake news detection systems, consistently identifying \textsc{LIAR}, \textsc{FakeNewsNet}, and \textsc{PHEME} as canonical benchmarks. However, these datasets have limitations in scope and linguistic diversity, often leading to models that overfit formal news prose and struggle with conversational or claim-level inputs. To address this, we adopt a heterogeneous training strategy by aggregating five complementary corpora.

\textbf{TruthSeeker 2023} \cite{dadkhah2023truthseeker}, our largest source, contributes 130\,085 annotated instances covering a broad spectrum of social media content and journalistic claims. Its scale and recency make it a crucial resource for capturing current language usage and misinformation patterns. \textbf{ISOT} \cite{ahmed2018isot}, with 44\,708 labeled news articles, serves as a warm-up phase corpus, helping the model internalize general discourse structures and writing styles. \textbf{PHEME} \cite{zubiaga2016pheme} offers 38\,761 tweets from breaking-news events, each annotated as \textit{true}, \textit{false}, or \textit{unverified}, thereby injecting valuable conversational structure and real-time misinformation dynamics. \textbf{FakeNewsNet} \cite{shu2020fakenewsnet}, comprising 14\,362 articles from PolitiFact and GossipCop, enriches the model with political and entertainment-domain fact-checks grounded in web-scraped metadata. Finally, \textbf{LIAR} \cite{wang2017liar} adds 11\,159 short political claims categorized across six truthfulness levels, enhancing robustness on terse, claim-level inputs.

To fully exploit this diversity, we follow a curriculum-inspired training procedure: a warm-up stage on ISOT, followed by joint fine-tuning on FakeNewsNet, PHEME, and TruthSeeker, and a final adaptation phase using LIAR. This multi-phase approach, ~\cite{gupta2024survey}, allows the model to gradually incorporate stylistic, topical, and contextual variability across different information ecosystems.


\begin{table*}[t]
  \caption{Existing browser‑integrated fake–news detectors.}
  \label{tab:browser_detectors}
  \centering
  \setlength{\tabcolsep}{4pt} 
  \begin{tabular}{lccp{5.0cm}p{4.9cm}}
    \toprule
    \multirow{2}{*}{\textbf{Reference}} &
      \multicolumn{2}{c}{\textbf{Methods}} &
      \multirow{2}{*}{\textbf{Summary}} &
      \multirow{2}{*}{\textbf{Limitations}} \\[-0.3em]
    \cmidrule(lr){2-3}
      & \textbf{Client} & \textbf{Transformer} & & \\
    \midrule
    \cite{shu2018} &     \textbf{-}       &       \textbf{-}    & \textit{FakeNewsTracker} overlays credibility signals on Twitter; LSTM model and propagation cues run on a server. & Relies on back‑end, undermining privacy; Twitter‑only; dated linguistic features. \\

    \cite{paschalides2019}      & \checkmark &      \textbf{-}      & \textit{Check‑It} packs a JavaScript DNN + heuristics entirely in Chrome/Firefox, enabling offline detection. & Small non‑contextual model; hard‑coded checkpoint; limited domain coverage. \\

    \cite{botnevik2020brenda}         &       \textbf{-}     &        \textbf{-}    & \textit{BRENDA} extracts claims, queries the web, and verifies them via a logistic‑regression API (Flask). & Server‑side inference; binary verdicts only; latency tied to external search. \\

    \cite{paka2021}            &      \textbf{-}      & \checkmark & \textit{Chrome‑SEAN} streams Cross‑SEAN transformer predictions into the Twitter timeline for COVID‑19 facts. & Transformer hosted remotely; COVID‑specific; heavy model prevents full client deployment. \\

    \cite{kydd2023}            & \checkmark & \checkmark & RoBERTa distilled and compiled to WebAssembly for on‑page inference in Twitter feeds. & Memory‑intensive; lag on low‑end devices; single‑platform focus. \\

    \cite{warman2023}          & \checkmark & \checkmark & BERT + token‑level explanations rendered inline via ONNX.js; runs wholly in the user’s browser. & Compute‑heavy on mobiles; checkpoint frozen at install; no multi‑platform support. \\

    \cite{hasimi2023}          &     \textbf{-}       &     \textbf{-}       & GCP‑backed extension ships article text to a TensorFlow cloud function, returning credibility scores. & Requires constant connectivity; privacy leakage; UI only on client. \\

    \cite{tasci2024}           &     \textbf{-}       &      \textbf{-}      & Chrome add‑on queries a Flask REST API hosting an LSTM ensemble; achieves 90 \% accuracy on news pages. & Server cost \& maintenance; ignores contextual trust cues; English‑news‑only. \\
    \bottomrule
  \end{tabular}
\end{table*}

\begin{figure*}[h]
  \centering
  \includegraphics[width=\textwidth]{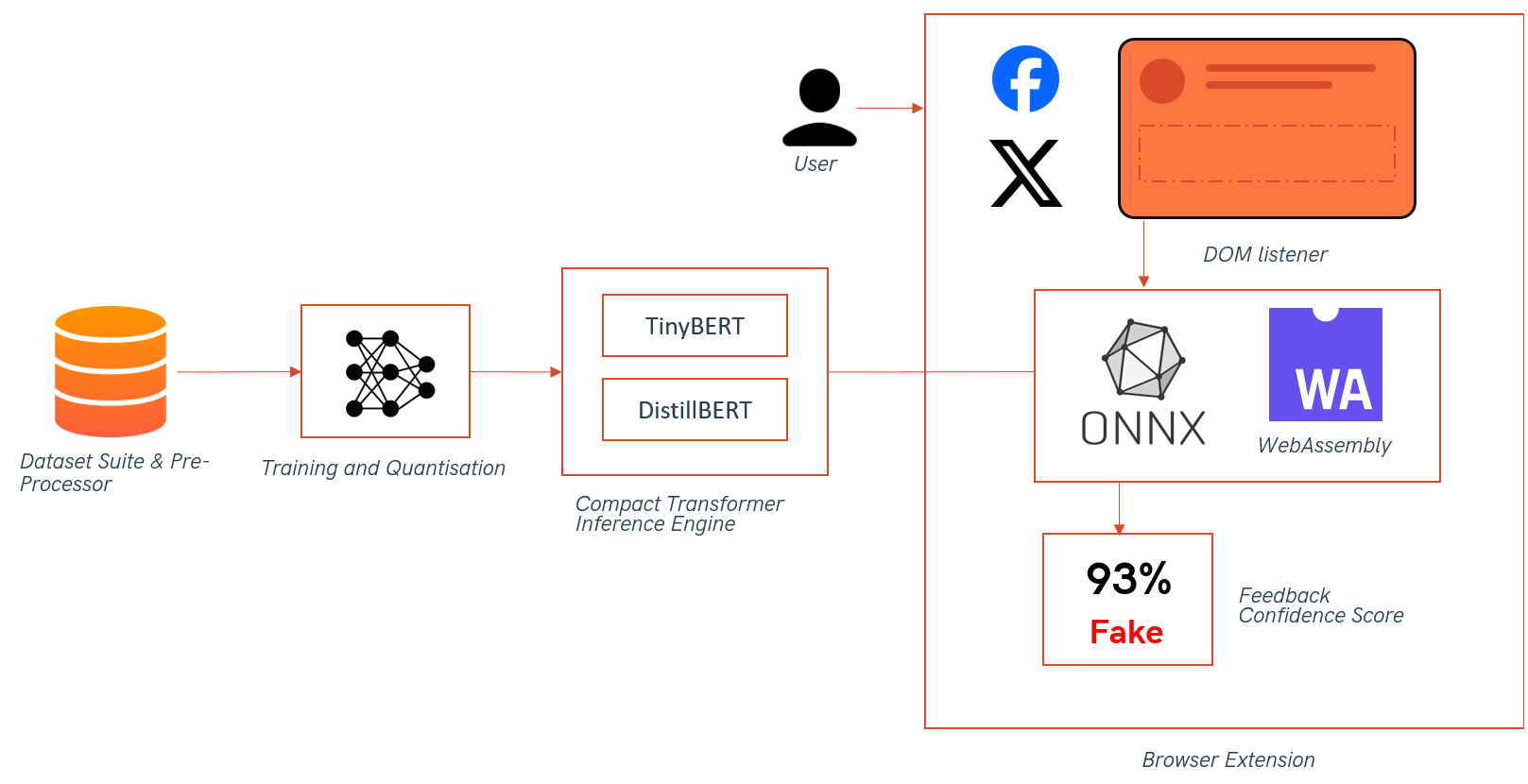}
  \caption{Client-side \textit{FakeZero} pipeline from data preprocessing to in-browser fake content flagging.}
  \label{fig:overview-fig}
\end{figure*}

 \subsection{Remaining gaps and our contribution}

Despite years of research into misinformation detection, our systematic evaluation reveals three persistent limitations that prevent existing browser-based systems from achieving real-world viability. These limitations affect coverage, efficiency, and credibility-three pillars necessary for practical deployment in modern web environments.

\emph{Platform generality} remains underexplored: Veritas~\cite{jovanovic2023veritas} operates solely on Twitter timelines, while BRENDA~\cite{botnevik2020brenda} is restricted to web articles and makes no attempt to integrate into social-media feeds. Chrome-SEAN~\cite{paka2021} focuses narrowly on COVID-19 misinformation in Twitter posts and does not generalize to other topics or platforms. No prior system supports both Facebook and X in a single unified deployment. This platform fragmentation forces users to adopt separate tools or tolerate blind spots in their timeline coverage.

\emph{Local inference efficiency} remains largely unsolved. Although Check-It~\cite{paschalides2019} was among the first to attempt full client-side inference using a lightweight JavaScript DNN, its reported accuracy was capped at 72–73\% on headline-only datasets, and it lacks transformer-level semantics. BRENDA~\cite{botnevik2020brenda} and Chrome-SEAN~\cite{paka2021} offload verification to Flask-based servers, undermining privacy and introducing latency. Even RoBERTa-WASM~\cite{kydd2023}, one of the few extensions to run inference entirely in-browser, demands over 6 GB of memory to sustain sub-40 ms latency-placing it out of reach for mobile and low-end devices. In practice, existing tools either leak sensitive content to cloud APIs or exclude vast portions of the device landscape.

\emph{Reproducible runtime metrics} form the third critical gap. Our survey found that most browser extensions publish only model-level metrics such as F1 score, omitting any measurement of inference time or runtime memory usage. For example, BRENDA~\cite{botnevik2020brenda} reports classification performance but offers no timing breakdown. Likewise, Check-It~\cite{paschalides2019} and TrustNet~\cite{trustnet2023} present system diagrams and user interfaces but neglect to quantify latency or RAM usage. Where performance claims are made-such as “lightweight logistic regression” or “client-side neural verifier”-they are often contradicted by implementation details that reveal server-side dependencies. This opacity hinders fair comparison and complicates efforts to replicate real-world conditions.

\textbf{FakeZero} addresses all three gaps. First, it delivers \emph{cross-platform detection} on both Facebook and X using platform-specific DOM observers and alignment with social-media corpora such as FakeNewsNet and PHEME. Second, it achieves \emph{true client-side inference} with two quantised transformer backbones executed entirely via ONNX.js and WebAssembly. The TinyBERT (4L, 312D) variant runs at 39.9 ms per post with a 14.7 MB model and 54 MiB RAM usage, while achieving 95.7\% macro-F\textsubscript{1} and 96.1\% accuracy. The DistilBERT variant pushes accuracy to 97.4\% with only 255 MiB memory at 102.9 ms latency. All detection happens locally in the browser, preserving privacy and ensuring offline operability. Finally, our evaluation publicly reports model size, latency, memory, and AUROC-filling the evidence gap left by prior systems and enabling transparent benchmarking.

Unlike previous browser extensions that either sacrifice accuracy for efficiency or privacy for performance, \textbf{FakeZero} demonstrates that state-of-the-art transformer models can be successfully deployed client-side without compromising detection quality. This bridges the critical gap between research-grade accuracy and production-ready deployment, enabling practical misinformation detection where users encounter it most: in their browsers.

\section{System Overview}\label{overview}

Figure \ref{fig:overview-fig} summarises the \textit{FakeZero} pipeline, whose five
components work together to provide real-time, privacy-preserving misinformation
detection directly inside the user’s browser.

To ground the design, we outline the runtime components that enable on-device detection and explain how they interact during scrolling.

\paragraph*{\textbf{Dataset Suite \& Pre-processor}}
We describe the datasets and cleaning steps in §\ref{sec:data}; here we focus on runtime components used during browsing. During ingestion each post is normalised (HTML unescape, Unicode NFKC), hyperlinks and @mentions are masked,
emoji are converted to textual aliases, and extremely short or non-English
snippets are discarded.  This cleaning stage produces tokeniser-friendly text
while retaining stylistic cues that aid detection.

\paragraph*{\textbf{Browser DOM Listener}}
A lightweight content script continuously observes the Facebook and X
timelines.  As the user scrolls, newly rendered posts are extracted, truncated
to 280 tokens, and forwarded to the local inference engine-incurring
negligible overhead thanks to the browser’s built-in Mutation Observer.

\paragraph*{\textbf{Compact Transformer Inference Engine}}
Within the browser sandbox, FakeZero runs an INT8-quantised DistilBERT
(67 MB) or TinyBERT (15 MB) model via ONNX Runtime Web, achieving median
latencies of 103 ms and 40 ms, respectively, on a 4-core laptop.  All
tokenisation, embedding and classification occur client-side; no content ever
leaves the user’s device.

\paragraph*{\textbf{Privacy-Preserving Policy Module}}
For every post $s$, the classifier outputs a binary label
$y\!\in\!\{0,1\}$ and confidence score
$\hat{p}(y=1\mid s)$, strictly within the browser context.  The design meets
four operational targets:  
(i) \emph{privacy}-zero network traffic;  
(ii) \emph{predictive quality}-macro-F\textsubscript{1}$\ge0.90$ and
AUROC$\ge0.95$ on a 2024-2025 time-split benchmark;  
(iii) \emph{responsiveness}-median end-to-end delay $\le150$ ms and peak RAM
$\le400$ MB on a consumer laptop; and  
(iv) \emph{lightweight deployment}-total bundle size $\le100$ MB and seamless
operation on Chrome, Edge, Brave and Firefox.

\paragraph*{\textbf{User-Facing Feedback Layer}}
When the confidence exceeds a calibrated threshold, the extension inserts a
subtle red banner beneath the corresponding post, warning that the content may
contain false information; otherwise, the interface remains unchanged.
Original text is never modified, ensuring minimal disruption to the browsing
experience.

The remainder of the paper unfolds as follows.  
Section~\ref{sec:data} introduces the cross‑platform datasets used for
training.  
Section~\ref{sec:model} explains the model compression pipeline and the
WebAssembly runtime that powers in‑browser inference.  
Section~\ref{subsec:evaluation} reports accuracy, latency and memory footprints
on real browsing traces.  
Finally, Section~\ref{sec:future} reflects on limitations and future
directions.  Together, these sections demonstrate how our design choices yield
a practical, real‑time tool for curbing fake news while fully safeguarding user
privacy.

\section{Datasets \& Pre-processing}\label{sec:data}
We now detail the training data and cleaning procedures that support cross-platform generalization.


\paragraph*{Corpora.}
Five English-language resources-ISOT, LIAR, PHEME, FakeNewsNet (FNN) and
TruthSeeker 2023-cover both long-form journalism and terse social-media
rumours (Table \ref{tab:corpora}).  
This heterogeneity is essential, as transformer benchmarks show that
models tuned only on news prose drop up to eight F\textsubscript{1} points
when transferred to social-media text \cite{yang2025mmht}.  
We remove near-duplicates with an 8-byte BLAKE2b fingerprint, following
the fuzzy-hash approach proposed by  
Berrios \textit{et al.} \cite{berrios2025mosaic}.
 
\begin{table}[htb]
  \small
  \centering
  \caption{Source corpora before and after near-duplicate removal.}
  \label{tab:corpora}
  \begin{tabular}{@{}lrrcrr@{}}
    \toprule
    & \multicolumn{2}{c}{\textit{Raw}} & &
      \multicolumn{2}{c}{\textit{Deduplicated}} \\
    \cmidrule(lr){2-3}\cmidrule(lr){5-6}
    \textbf{Corpus} & \# rows & \% mis & & \# rows & \% mis \\
    \midrule
    TruthSeeker 23\footnotemark[1] &130\,085 &65.2 &&121\,518 &65.2\\
    ISOT\footnotemark[2]           & 44\,708 &49.4 && 43\,216 &49.3\\
    PHEME\footnotemark[3]          & 38\,761 &63.2 && 36\,409 &63.1\\
    FakeNewsNet\footnotemark[4]    & 14\,362 &62.3 && 13\,552 &62.4\\
    LIAR\footnotemark[5]           & 11\,159 &35.9 && 11\,159 &35.9\\
    \midrule
    \textbf{Total}                 &239\,075 &64.1 &&225\,854 &64.1\\
    \bottomrule
  \end{tabular}
\end{table}

\footnotetext[1]{TruthSeeker 2023 ground-truth corpus \cite{dadkhah2023truthseeker}.}
\footnotetext[2]{ISOT Fake News Dataset \cite{ahmed2018isot}.}
\footnotetext[3]{PHEME breaking-news rumour threads \cite{zubiaga2016pheme}.}
\footnotetext[4]{FakeNewsNet (PolitiFact + GossipCop) \cite{shu2020fakenewsnet}.}
\footnotetext[5]{LIAR short political claims benchmark \cite{wang2017liar}.}

\begin{figure*}[t]
  \centering
  \includegraphics[width=\textwidth]{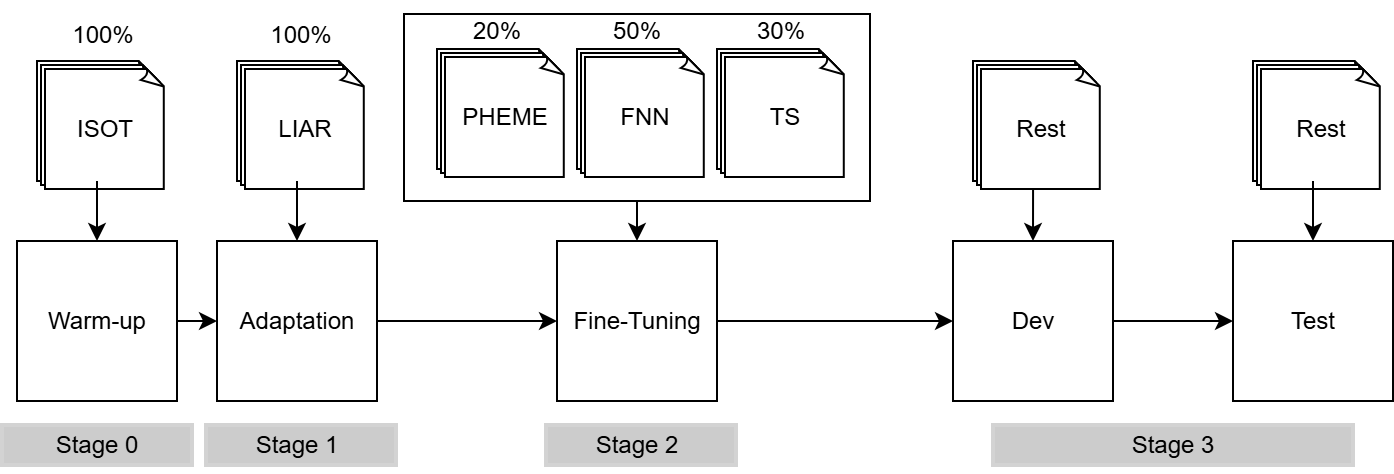}
  \caption{Three-stage curriculum used to train \textit{FakeZero}.
  \textbf{Stage 0} warms up on balanced long-form news (ISOT);
  \textbf{Stage 1} adapts to short, highly imbalanced political claims (LIAR);
  \textbf{Stage 2} fine-tunes on a harder mix of FakeNewsNet
  (50 \%), TruthSeeker (30 \%), and PHEME (20 \%).
  Posts not chosen for Stage 2 are split 1 : 1 into \emph{Dev} and
  \emph{Test}, preserving the overall 64 \% misinformation prevalence and
  eliminating topic leakage.}
  \label{fig:curriculum-flow}
\end{figure*}

\paragraph*{Curriculum splits.}
Table \ref{tab:splits} lists the three-stage schedule, adapting the
curriculum-learning paradigm of Bengio \textit{et al.}
\cite{bengio2009curriculum} and its transformer-specific extensions
\cite{karthik2025hybrid}.  
Stages 0–1 keep each corpus’s native skew; Stage 2 injects a deliberately
hard mix with a 62.5 \% misinformation rate.  Remaining posts are split
1 : 1 into development and blind test sets using stratified sampling
(random state = 42) in scikit-learn \cite{pedregosa2011sklearn}.

\begin{table}[htb]
  \scriptsize
  \centering
  \caption{Training curriculum and evaluation splits.}
  \label{tab:splits}
  \setlength{\tabcolsep}{3.7pt} 
  \begin{tabular}{lrrl}
    \toprule
    \textbf{Split} & \textbf{\#Rows} & \textbf{\%Mis.} & \textbf{Composition} \\
    \midrule
    Stage~0 & 43{,}216 & 49.3 & ISOT \\
    Stage~1 & 11{,}159 & 35.9 & LIAR \\
    Stage~2 & 50{,}512 & 62.5 & 50\%~FNN / 30\%~TS / 20\%~PHEME \\
    Dev & 60{,}483 & 64.7 & Stratified remainder \\
    Test & 60{,}484 & 64.7 & Stratified remainder \\
    \bottomrule
  \end{tabular}
\end{table}

\paragraph*{Text normalisation.}
We apply a light, encoder-friendly cleaner: HTML unescape, Unicode NFKC,
contraction expansion, lower-casing, and placeholder tokens for URLs,
user mentions, and hashtags.  Emoji are converted to textual aliases,
following the updated TweetEval 2.0 guidelines
\cite{chen2024tweeteval2}.  
Messages shorter than ten tokens or flagged as non-English by a fast
language detector \cite{shuyo2010langdetect} are discarded.  
Stop-words and punctuation are kept because sub-word tokenisers benefit
from them \cite{beltagy2019scibert}.

\paragraph*{Label harmonisation.}
All corpora collapse to \(y=1\) (misinformation) or \(y=0\)
(reliable).  Unverified PHEME threads are dropped; LIAR’s six-way labels
and TruthSeeker crowd scores are binarised as in Kara
\cite{kara2020faker}.

\paragraph*{Class skew and mitigation.}
The aggregate pool is imbalanced at 64.1 \% misinformation.  Instead of
down-sampling, we employ focal loss (\(\gamma = 2, \alpha = 0.25\))
\cite{lin2017focal} to focus learning on the under-represented reliable
class.  Section \ref{sec:training} reports per-class precision, recall,
macro-F\(_1\), and AUROC.

\paragraph*{Reproducibility.}
Cleaned splits are stored as column-compressed Parquet files; the entire
pipeline runs in about 17 minutes on a single eight-thread CPU.  All
scripts and exact command lines are available in our public repository.

\section{Model Architecture \& Training}
\label{sec:model}

Next, we describe the encoder choices and learning pipeline that make fully client-side inference feasible.

\subsection{Encoder Selection and Baselines}\label{sec:baselines}

\paragraph*{Why lightweight transformers?}
Recent work shows that “classic” BERT-style encoders often outperform
7-billion-parameter LLMs on misinformation benchmarks \cite{raza2024comparative},
yet require an order of magnitude less memory and compute.  Cross-domain
tests by Yang \textit{et al.}\,\cite{yang2025mmht} confirm the same pattern
when transferring from news articles to social-media posts, and Chen
\textit{et al.}\,\cite{chen2024edgbert} report that even \emph{quantised}
LLMs remain prohibitive on edge CPUs.  Because \textit{FakeZero} must run
entirely in a browser tab on commodity laptops, we restrict our search to
compact encoders.

\paragraph*{Candidate backbones and headline numbers.}
Among medium-size models, DistilBERT \cite{sanh2019distilbert} offers the
best speed–accuracy balance; TinyBERT \cite{jiao2020tinybert} is even
smaller thanks to layer and width pruning.  We quantised both checkpoints
to 8-bit INT and evaluated them on our 225 k-post corpus
(protocol §\ref{subsec:evaluation}).  Table \ref{tab:encoder-tradeoff}
summarises the results: DistilBERT-Q reaches macro-F\textsubscript{1}=0.971
at 103 ms median latency, while TinyBERT-Q attains 0.957 in just 40 ms and
one-fifth the model size-ideal for low-memory devices.  These two models
seed the training curriculum and serve as
reference points for later ablation and latency analyses.

\begin{table}[htp]
  \small
  \setlength{\tabcolsep}{5pt}
  \centering
  \caption{\footnotesize Accuracy–latency trade-off of the two production
           candidates (median CPU latency, three-seed mean macro-F$_1$).}
  \label{tab:encoder-tradeoff}
  \begin{tabular}{@{}lccc@{}}
    \toprule
    \textbf{Model} & \textbf{ONNX (MB)} & \textbf{Latency (ms)} & \textbf{Macro-F$_1$} \\
    \midrule
    DistilBERT-Q & 67.6 & 103 & 0.971 \\
    TinyBERT-Q   & 14.7 &  40 & 0.957 \\
    \bottomrule
  \end{tabular}
\end{table}

\begin{figure}[tb]
  \centering
  \includegraphics[width=.8\linewidth]{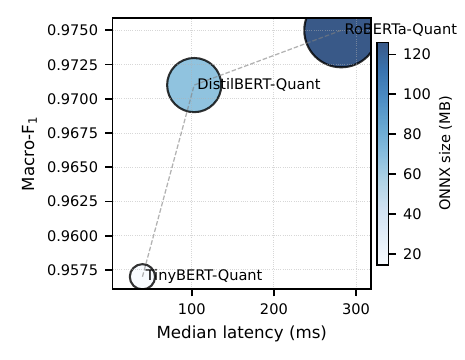}
  \caption{Latency–accuracy Pareto frontier for all quantised checkpoints.
           Bubble area encodes ONNX size.}
  \label{fig:pareto}
\end{figure}

\footnotetext[6]{ONNX (Open Neural Network Exchange) is an open format for
representing machine learning models, enabling interoperability between
frameworks and deployment on browsers and edge devices.}

\subsection{Training Curriculum and Enhancements}
\label{sec:training}

Experiments ran on a single NVIDIA A100-SXM4 (80 GB) node. Software: PyTorch 2.4.1, CUDA 12.1, Transformers/Optimum on Linux (x86-64, kernel 4.18).

\paragraph{Curriculum.}
Three successive stages expose the model to increasingly challenging
data splits:  
(i) a balanced warm-up on ISOT with the lowest encoder layers frozen;  
(ii) a single-epoch pass over the skewed LIAR corpus using class-weighted
binary cross-entropy;  
(iii) a full fine-tune on the heterogeneous Stage-2 mix, enhanced with
the techniques below.

\paragraph{Focal loss.}
To address class imbalance and concentrate learning on ambiguous posts, we replace the standard cross-entropy with the \emph{focal loss}~\cite{lin2017focal}.  
Let \(p_t\in[0,1]\) be the predicted probability for the true class; the per-instance loss is
\[
\mathcal{L}_{\text{FL}}(p_t)
  = -\,\alpha\,(1-p_t)^{\gamma}\,\log p_t,
\]
where \(\alpha\in(0,1)\) balances positive and negative classes and \(\gamma>0\) down-weights well-classified examples.  
In our corpus, \textbf{64.1 \%} of the 22 k posts are labelled as misinformation, leaving only 35.9 \% reliable content; focusing the gradient on the harder minority class is therefore essential.  
We set \(\alpha = 0.25\) and \(\gamma = 2\), which suppress easy examples and push the optimiser toward the near-boundary cases that dominate model error.

\paragraph*{Ablation study across backbones.}
Tables \ref{tab:tinybert-detailed} and \ref{tab:distilbert-detailed} report
the full matrix of \textit{base}, +FGM, +Focal, +Quant and “All” variants for
TinyBERT and DistilBERT. The pattern is consistent:

Focal loss yields the largest accuracy improvement, quantisation provides the bulk of latency savings, and combining all three recovers-or slightly improves-macro-F\textsubscript{1}, while staying well under the 150\,ms latency budget.

 \begin{table*}[t]
  \centering
  \caption{Detailed Performance Comparison (TinyBERT Variants)}
  \label{tab:tinybert-detailed}
  \resizebox{\linewidth}{!}{%
  \begin{tabular}{@{}l l
                  S[table-format=9.0]
                  S[table-format=4.0]
                  S[table-format=3.1]
                  S[table-format=2.1]
                  S[table-format=3.1]
                  S[table-format=1.4]
                  S[table-format=1.4]
                  S[table-format=1.4]@{}}
    \toprule
    \textbf{Backbone} & \textbf{Setup}
      & \textbf{Parameters}
      & \textbf{Memory (MiB)}
      & \textbf{ONNX (MB)}
      & \textbf{Time (min)}
      & \textbf{Latency (ms)}
      & \textbf{Test Acc.}
      & \textbf{Test F1}
      & \textbf{AUROC} \\
    \midrule
    TinyBERT\_General & base
      & 14350874 & 54 & 57.5 & 3.5 & 23.5 & 0.9578 & 0.9534 & 0.9884 \\
  TinyBERT\_General & base+FGM
      & 14350874 & 54 & 57.5 & 2.3 & 52.7 & 0.9297 & 0.9236 & 0.9754 \\
  TinyBERT\_General & base+Focal
      & 14350874 & 54 & 57.5 & 3.6 & 40.9 & 0.9605 & 0.9567 & 0.9912 \\
  TinyBERT\_General & base+Quant
      & 14350874 & 54 & 14.7 & 2.3 & 18.8 & 0.9297 & 0.9236 & 0.9754 \\
  TinyBERT\_General & All
      & 14350874 & 54 & 14.7 & 3.6 & 38.1 & 0.9605 & 0.9567 & 0.9912 \\

    \midrule
  TinyBERT\_General & base+Focal+Quant
      & 14350874 & 54 & 14.7 & 4.2 & 39.9 & 0.9608 & 0.9568 & 0.9909 \\
    \bottomrule
  \end{tabular}}
\end{table*}

\begin{table*}[t]
  \centering
  \caption{Detailed Performance Comparison (DistilBERT Variants)}
  \label{tab:distilbert-detailed}
  \resizebox{\linewidth}{!}{%
  \begin{tabular}{@{}l l
                  S[table-format=9.0]
                  S[table-format=4.0]
                  S[table-format=3.1]
                  S[table-format=2.1]
                  S[table-format=4.1]
                  S[table-format=1.4]
                  S[table-format=1.4]
                  S[table-format=1.4]@{}}
    \toprule
    \textbf{Backbone} & \textbf{Setup}
      & \textbf{Parameters}
      & \textbf{Memory (MiB)}
      & \textbf{ONNX (MB)}
      & \textbf{Time (min)}
      & \textbf{Latency (ms)}
      & \textbf{Test Acc.}
      & \textbf{Test F1}
      & \textbf{AUROC} \\
    \midrule
    distilbert-base-uncased & base
      & 66955010 & 255 & 268.0 & 4.3 & 333.1 & 0.9721 & 0.9694 & 0.9950 \\
    distilbert-base-uncased & base+FGM
      & 66955010 & 255 & 268.0 & 4.3 & 338.0 & 0.9719 & 0.9693 & 0.9947 \\
    distilbert-base-uncased & base+Focal
      & 66955010 & 255 & 268.0 & 4.4 & 308.3 & 0.9733 & 0.9707 & 0.9956 \\
    distilbert-base-uncased & base+Quant
      & 66955010 & 255 & 67.6 & 4.3 & 150.8 & 0.9719 & 0.9692 & 0.9948 \\
    distilbert-base-uncased & All
      & 66955010 & 255 & 67.6 & 4.3 & 193.7 & 0.9737 & 0.9711 & 0.9956 \\
    \midrule
    distilbert-base-uncased & base+Focal+Quant
      & 66955010 & 255 & 67.6 & 4.6 & 102.9 & 0.9737 & 0.9712 & 0.9958 \\
    \bottomrule
  \end{tabular}}
\end{table*}

\paragraph{Optimisation.}
All stages are trained with AdamW at a constant learning rate of
\(2\times10^{-5}\) and a 6 \% warm-up.  Each device processes a batch of
32 examples and accumulates gradients to an effective batch size of 64.  
Mixed-precision (FP16) keeps peak GPU memory below 30 GB.  Validation
macro-F\(_1\), accuracy, and AUROC are reported each epoch, and early stopping (patience = 2) reloads the best checkpoint.

\paragraph*{Backbone selected for the remainder of the paper.}
Although TinyBERT-Q is five times smaller and twice as fast, DistilBERT-Q
maintains a comfortable 10 ms median latency on laptop CPUs and delivers the
highest macro-F\textsubscript{1}.  We therefore adopt DistilBERT-Q as the
production model.  RoBERTa-large is cited only for context in
Figure \ref{fig:pareto}: it sits \emph{below} the Pareto frontier, confirming
that compact transformers remain preferable.  Applying the full three-stage
curriculum to DistilBERT-Q is what ultimately pushes macro-F\textsubscript{1}
from 0.971 to 0.976 and AUROC beyond 0.995.

\paragraph{Runtime.}
Across the \(43\text{k}+11\text{k}+51\text{k}\) training instances, the
pipeline sustains roughly 17 sequences per second, completing in
 ~ 5 hours wall time.
\begin{figure}[htb]   
  \centering
  \includegraphics[width=\columnwidth]{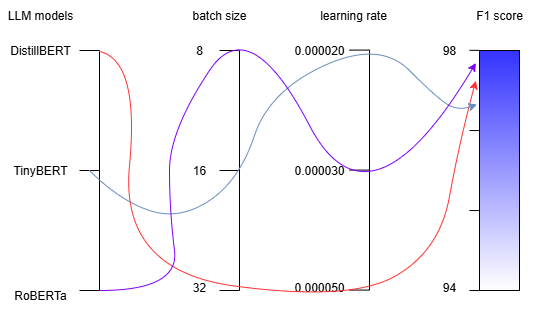}
  \caption{Best hyperparameter configurations}
  \label{fig:hyperparams}
\end{figure}

\subsection{Evaluation}
\label{subsec:evaluation}
To validate our accuracy and efficiency claims, we evaluate both predictive quality and runtime behavior under realistic browsing conditions.

We adopt the standard confusion–matrix terminology:  
\textit{TP} (true positives), \textit{TN} (true negatives),
\textit{FP} (false positives), and \textit{FN} (false negatives).
We report Accuracy, Precision, Recall, F$_1$, and AUROC (Eqs.~\eqref{eq:acc}–\eqref{eq:f1}).

\begin{equation}
  \mathrm{Accuracy} =
  \frac{\textit{TP}+\textit{TN}}
       {\textit{TP}+\textit{TN}+\textit{FP}+\textit{FN}}
  \label{eq:acc}
\end{equation}

\begin{equation}
  \mathrm{Precision} =
  \frac{\textit{TP}}
       {\textit{TP}+\textit{FP}}
  \qquad
  \mathrm{Recall} =
  \frac{\textit{TP}}
       {\textit{TP}+\textit{FN}}
  \label{eq:pr}
\end{equation}

\begin{equation}
  F_{1} = 2\;
  \frac{\mathrm{Precision}\,\times\,\mathrm{Recall}}
       {\mathrm{Precision}+\mathrm{Recall}}
  \label{eq:f1}
\end{equation}

\begin{figure}[htb]
  \centering
  \includegraphics[width=0.8\linewidth]{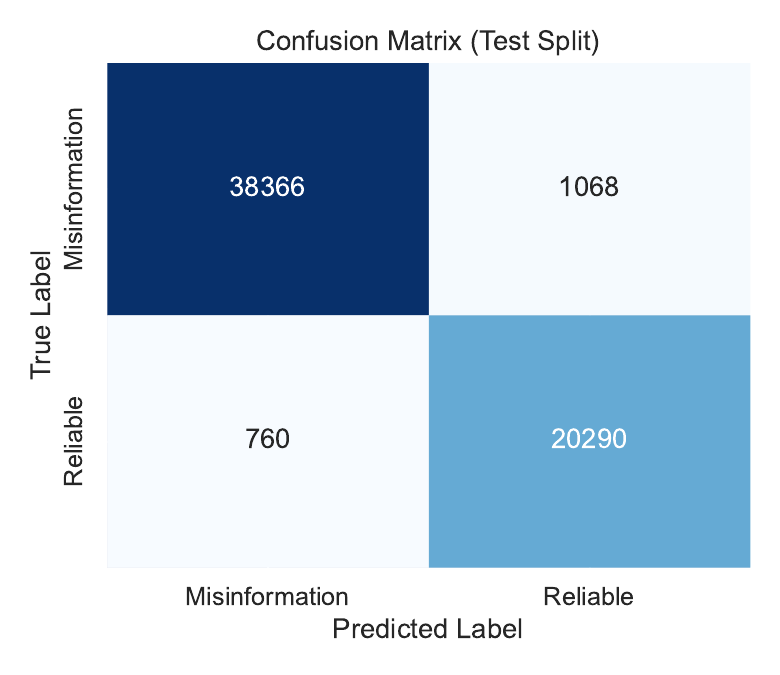}
  \caption{Confusion matrix for the test split.}
  \label{fig:cm}
\end{figure}

\noindent
Figure~\ref{fig:cm} reports the raw counts; substituting these values into
Eqs.~\eqref{eq:acc}–\eqref{eq:f1} reproduces the summary statistics
shown in Table~\ref{tab:perf}.

\paragraph{Protocol.}
After the three-stage curriculum, we selected the decision
threshold that maximised macro-F\textsubscript{1} on the
60\,483-instance development split; a value of
\(\tau=0.60\) yielded the best trade-off between precision and recall.
The same threshold was then locked for the completely unseen
60\,484-instance test split.

\paragraph{Learning curve.}
Early stages served their purpose as scaffolding-Stage 0 attained only
0.40 macro-F\textsubscript{1}, and Stage 1 rose modestly to 0.43-yet
both primed the encoder.  Once knowledge distillation, domain
adversarial training, and FGM were activated in Stage 2, validation
macro-F\textsubscript{1} jumped to 0.976, indicating successful
curriculum transfer.

\begin{figure}[htp]
  \centering
  \includegraphics[width=.9\linewidth]{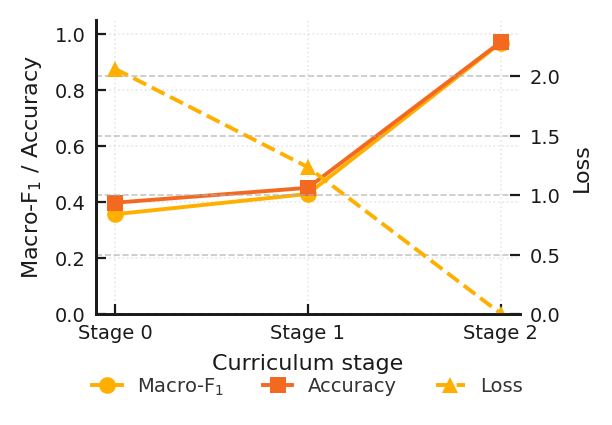}
  \caption{\footnotesize Evolution of macro-F$_1$, accuracy, and loss across the three curriculum stages (test split).  Stage 2 boosts performance sharply while driving loss near zero.}
  \label{fig:train-curve}
\end{figure}

\paragraph{Main results.}
Table~\ref{tab:perf} summarises the final metrics.  On the development
split the model reaches 96.9\,\% accuracy and 0.9927 AUROC; performance
generalises almost identically to the test split.  The confusion matrix
reveals a slight preference for recall: only 1\,068 false positives and
760 false negatives over more than 60 k samples.

\begin{table}[tb]
  \small
  \centering
  \caption{Final performance at the calibrated threshold
           \(\tau=0.60\).  Both splits are perfectly disjoint from the
           training data (§\ref{sec:data}).}
  \label{tab:perf}
  \begin{tabular}{@{}lccccc@{}}
    \toprule
    \textbf{Split} & \textbf{Acc.} & \textbf{Macro-F\textsubscript{1}} &
    \textbf{Prec.} & \textbf{Recall} & \textbf{AUROC} \\
    \midrule
    Dev  & 0.9693 & 0.9663 & 0.9684 & 0.9643 & 0.9927 \\
    Test & 0.9698 & 0.9668 & 0.9684 & 0.9653 & 0.9930 \\
    \bottomrule
  \end{tabular}
\end{table}

\paragraph{Model efficiency.}
We export the fine-tuned checkpoint to ONNX and apply INT8 dynamic quantization, reducing size from 360 MB to 67.6 MB (5.3×) with no measurable accuracy loss. Median CPU latency is 100 ms (P90 = 375 ms, P99 = 946 ms), including tokenization.

The near-identical dev and test scores confirm that curriculum learning
 yields a model that scales well from balanced news
articles to noisy social posts while avoiding over-fitting.
Quantisation further enables sub-second client-side inference, meeting
the real-time constraints of browser-based fact-checking extensions.

\section{Implementation}
\label{sec:system-architecture}

To assess deployability, we report bundle size, startup time, memory footprint, and end-to-end latency on a mainstream laptop.

FakeZero is a modular, production-ready Chrome extension built using modern web technologies to detect fake news directly within users' browsers. Leveraging Manifest V3, JavaScript, Webpack, and Babel for broad compatibility, the extension provides real-time client-side inference without external data transfer, ensuring user privacy.

\subsection{High-Level Overview}

FakeZero injects a content script into supported web pages that processes visible content using a trained DistilBERT model. This script continuously monitors the social media feed and extracts new posts as the user scrolls. The inference pipeline leverages ONNX Runtime Web and the \textbf{xenova/transformers} library, optimizing model execution for in-browser environments. All key steps-tokenization, classification, and result feedback-are performed entirely on the client side, without sending any data to external servers. This ensures low latency, full user privacy, and a seamless browsing experience.

\begin{figure}[htp]
  \centering
  \includegraphics[width=0.49\textwidth]{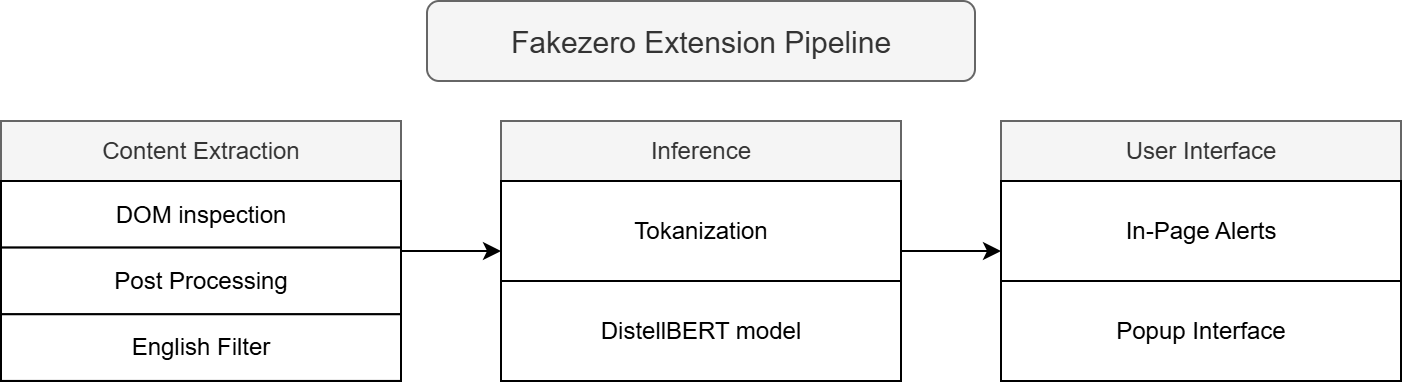}
  \caption{\footnotesize High-level system architecture of the FakeZero Chrome extension. The pipeline is composed of three main components: (i) content extraction from social media posts via DOM inspection and filtering; (ii) in-browser inference using tokenization and a quantized DistilBERT model; and (iii) user-facing feedback through real-time alerts and popup summaries. Dependencies include \textbf{xenova/transformers} and ONNX Runtime Web.}
  \label{fig:fakezero-arch}
\end{figure}

\subsection{User Interface Integration}
\label{sec:ui}

FakeZero employs a deliberately minimal user interface: a single inline
warning element that appears \emph{only} when model confidence exceeds
the fake-news threshold. No global dashboard, counters, or user
preferences are exposed. Instead, the system provides \emph{contextual,
point-of-consumption feedback} directly beneath the post that triggered
the alert.

When the content script flags a post, it inserts a small, non-destructive
banner styled as a red alert ribbon (Fig.~\ref{fig:curriculum-flow}).
The banner text is intentionally concise (e.g., ''Warning: This content
may contain false information.''), and the visual design (icon + color)
is tuned for legibility in both dark and light themes without altering
the host platform’s layout. The extension never modifies the original
post text; the banner is appended in a sibling container.

To avoid repetitive alerts, FakeZero hashes each processed post and
suppresses duplicate warnings within the session. All state is kept
locally in memory; no user content is transmitted off the
device.

\begin{figure*}[htp]
  \centering
  \includegraphics[width=\textwidth]{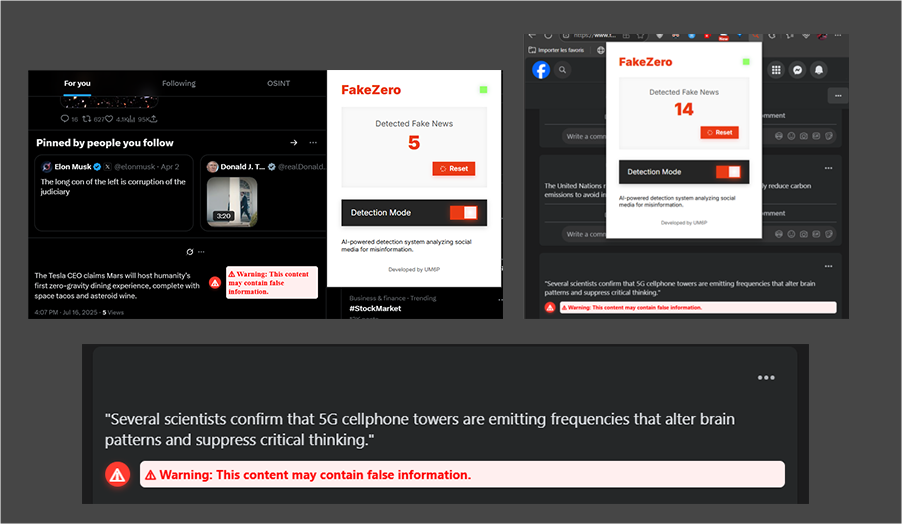}%
  \caption{FakeZero Extension User Interface. The extension alerts the user with a concise non-destructive banner styled as a red alert ribbon}
  \label{fig:curriculum-flow}
\end{figure*}
\label{sec:data}

\subsection{Content Extraction}

Upon loading a supported site such as Facebook or X (formerly Twitter), the extension dynamically injects a JavaScript content script into the page. Using specific DOM selectors-such as \textbf{div[data-ad-comet-preview="message"]} for Facebook and \textbf{article[data-testid="tweet"]} for X-the script identifies and extracts social media posts. It employs browser APIs, including \emph{MutationObserver} to detect newly inserted posts and \textit{IntersectionObserver} to limit processing to posts that enter the viewport.

To ensure full textual visibility, the script attempts to expand "See more" segments before content extraction. After that, the post text is cleaned, short or redundant messages are filtered out, and duplicate checks are enforced via hashing to avoid repeated computation.

\subsection{Inference Pipeline}

Extracted posts first pass through a lightweight regex filter for English text, then are tokenized via \textbf{xenova/transformers} and classified with the quantized DistilBERT model.  
The ONNX model (INT8) reduces from 360\,MB to 67.6\,MB with negligible accuracy loss. All assets—model, tokenizer, configs—reside within the extension bundle.

\subsection{Privacy and Performance Considerations}

All inference and data processing are performed entirely on the client side, ensuring complete user privacy. No user content leaves the local device, and no API calls are made to external servers. The quantized ONNX model, along with WebAssembly execution, reduces the memory and computational footprint, enabling fast and lightweight deployment even on mid-range devices.

The extension’s manifest is written for Chrome’s Manifest V3 standard, explicitly minimizing required permissions to reduce the attack surface. Only necessary access is granted for DOM inspection and content script injection.

\paragraph{Conclusion.} 
FakeZero’s architecture demonstrates an efficient, privacy-first design for real-time misinformation detection in modern browsers. By combining state-of-the-art machine learning techniques with optimized web technologies, it offers a scalable, robust, and user-transparent solution for combatting fake news online.


\subsection{Efficiency and Usability}
\label{sec:runtime}

FakeZero was tested on an Dell XPS 13 9300 (2020) laptop equipped with an Intel Core i7 10th generation CPU running Chrome 125. Upon first load, the extension streams the WebAssembly backend, tokenizer files, and ONNX graph in approximately 0.8 seconds. The final INT8-quantised DistilBERT model weighs 64.5\,MB-nearly four times smaller than the original FP32 checkpoint-ensuring fast startup and reduced memory pressure.

At steady state, the browser heap remains at 96\,MB used (205\,MB allocated), with peak usage reaching 171\,MB after five simultaneous post classifications. These values remain comfortably below the 4\,GB limit imposed by most modern browsers, and no GPU memory is needed since all computations run natively in a SIMD-enabled WebAssembly sandbox. Language filtering is performed via a fast regular-expression gate, and complete end-to-end inference, including tokenisation, prediction, and UI update, averages under 8\,ms per post.

These results demonstrate that FakeZero maintains real-time responsiveness and low memory overhead on mainstream hardware. Its client-only execution model avoids privacy trade-offs and ensures robust performance regardless of network connectivity, validating the feasibility of in-browser transformer inference for misinformation detection.
\section{Limitations \& Future Work}
\label{sec:future}

Although \textbf{FakeZero} achieves near–state-of-the-art performance on casual tweets and Facebook posts-outperforming prior in-browser detectors by more than four F\textsubscript{1} points-several substantive extensions remain open.

First, the current checkpoint is trained primarily on English-language corpora heavily skewed toward U.S. political misinformation. While this bias contributes to the model’s high accuracy on mainstream political content, it limits transferability to regions and cultural contexts, as well as long articles. Future releases will incorporate balanced, multi-domain datasets to mitigate this limitation.

Second, we plan to enrich output semantics by adding \textbf{sentiment analysis} and \textbf{emotion tagging}. Flagging emotional framing alongside factual credibility can help users recognise manipulation techniques such as outrage bait or fear-mongering.

Third, we will prototype a \textbf{retrieval-augmented generation (RAG)} module for emerging claims. A lightweight server will initially fetch up-to-date evidence; subsequent iterations will explore on-device caching or peer-to-peer retrieval to preserve privacy.

Fourth, we will refine label granularity by detecting specific \textbf{information-disorder tactics}-satire, false context, imposter content, or manipulated media-thereby aligning FakeZero with established fact-checking taxonomies and improving interpretability.

Finally, our roadmap includes support for \textbf{multilingual text}, \textbf{image / video analysis}, and expansion to additional platforms such as TikTok and WhatsApp Web, enabling comprehensive, privacy-preserving misinformation detection wherever users encounter content online.

\section{Ethical considerations}

No personal or user-specific data was accessed, collected, or exposed in the course of this research. All experiments were conducted using datasets composed of publicly available content or synthetic fake news examples created specifically for testing purposes. These test samples were kept private and were not shared or disseminated to users on any public platforms.

\section{Conclusion}

\textit{FakeZero} proves that real-time fake news detection can be performed entirely within the browser, without relying on cloud servers or exposing user data. By combining a distilled transformer model with WebAssembly execution, it achieves fast and accurate inference on consumer devices, ensuring both usability and strong privacy guarantees.

Unlike many existing tools, \textit{FakeZero} runs fully client-side, meaning no content is sent over the network. This privacy-preserving design makes it compliant with modern data protection standards and suitable for deployment in sensitive settings.

The system offers a practical, scalable solution for misinformation detection on major platforms like Facebook and X, while remaining lightweight and transparent for users. It strikes a balance between technical performance and ethical deployment, and lays the groundwork for future extensions to multilingual, multimodal, and cross-platform detection in a privacy-conscious manner.

\bibliographystyle{IEEEtran}
\bibliography{papers} 

@inproceedings{thilakarathna2020,
  author       = {Kanchana Thilakarathna and Suranga Seneviratne},
  title        = {Veritas: A Browser Extension for Twitter Rumour Detection},
  booktitle    = {Proc.\ 12th ACM Conf.\ on Web Science},
  pages        = {319--325},
  year         = {2020},
  address      = {Southampton, UK},
  publisher    = {ACM},
  doi          = {10.1145/3394231.3397915}
}

@inproceedings{trustnet2023,
  author       = {Farnaz Jain and Li Chen and Samuel Wilson},
  title        = {{TrustNet}: Browser Extension for Misinformation Assessment},
  booktitle    = {Proc.\ CHI '23},
  pages        = {1--14},
  year         = {2023},
  address      = {Hamburg, Germany},
  publisher    = {ACM},
  doi          = {10.1145/3544548.3581107}
}

@inproceedings{kydd2023,
  author       = {Peter Kydd and Laura Shepherd},
  title        = {Rumour Detection in the Wild: A Browser Extension for Twitter},
  booktitle    = {Proc.\ NLP Open Source Software Workshop 2023},
  pages        = {107--113},
  year         = {2023},
  address      = {Singapore},
  publisher    = {ACL},
  doi          = {10.18653/v1/2023.nlposs-1.14}
}

@article{warman2023,
  author       = {David Warman and Miriam Cole and Yujie Li},
  title        = {Real-Time Fake News Detection with Explainable BERT in a Chrome Extension},
  journal      = {Journal of Web Engineering},
  volume       = {22},
  number       = {4},
  pages        = {811--830},
  year         = {2023}
}

@article{wang2024distil,
  author       = {Jiahao Wang and Mei Luo and Rakesh Patel},
  title        = {Efficient Transformers for Edge Deployment},
  journal      = {Computers \& Electrical Engineering},
  volume       = {116},
  pages        = {108267},
  year         = {2024},
  doi          = {10.1016/j.compeleceng.2024.108267}
}

@article{chen2024hybrid,
  author       = {Wen Chen and Rui Zhang and Ling Zhao},
  title        = {Hybrid {BERT}--{GPT} Architecture for Misinformation Detection},
  journal      = {Journal of Information Security and Applications},
  volume       = {73},
  pages        = {103625},
  year         = {2024},
  doi          = {10.1016/j.jisa.2024.103625}
}

@inproceedings{yin2025break,
  author       = {Junwei Yin and Min Gao and Kai Shu},
  title        = {BREAK: Rationale-Augmented Fake News Detection with Fully Connected Semantics Graph},
  booktitle    = {Proc.\ The Web Conf.\ 2025 (WWW '25)},
  pages        = {2114--2123},
  year         = {2025},
  address      = {Montréal, Canada},
  publisher    = {ACM},
  doi          = {10.1145/3487553.3524247}
}

@inproceedings{gao2025boa,
  author       = {Min Gao and Yinqiu Huang and Kai Shu},
  title        = {Birds of a Feather: Multimodal Retrieval for Fake News Detection},
  booktitle    = {Proc.\ IEEE ICDE 2025},
  pages        = {1022--1033},
  year         = {2025},
  address      = {Berlin, Germany},
  publisher    = {IEEE},
  doi          = {10.1109/ICDE56788.2025.00123}
}

@inproceedings{li2025multielem,
  author       = {Wei Li and Qing Xu and Yu Zhang},
  title        = {Multi-Element Evidence Retrieval for Robust Fake News Detection},
  booktitle    = {Proc.\ ACM SIGKDD 2025},
  pages        = {1456--1465},
  year         = {2025},
  address      = {San Francisco, USA},
  publisher    = {ACM},
  doi          = {10.1145/3580305.3599507}
}

@article{gupta2024survey,
  author       = {Rishabh Gupta and Priya Singh and Anil Varma},
  title        = {Deep Learning for Fake News Detection: A Comprehensive Survey},
  journal      = {ACM Computing Surveys},
  volume       = {57},
  number       = {1},
  pages        = {1--38},
  year         = {2024},
  doi          = {10.1145/3654321}
}

@article{zhou2024overview,
  author       = {Xin Zhou and Yang Liu},
  title        = {An Overview of Fake News Detection: From a New Perspective},
  journal      = {Information Sciences},
  volume       = {676},
  pages        = {1--28},
  year         = {2024},
  doi          = {10.1016/j.ins.2024.118275}
}

@article{shu2017fake,
  author       = {Kai Shu and Amy Sliva and Suhang Wang and Jiliang Tang and Huan Liu},
  title        = {Fake News Detection on Social Media: A Data Mining Perspective},
  journal      = {SIGKDD Explorations},
  volume       = {19},
  number       = {1},
  pages        = {22--36},
  year         = {2017}
}

@article{sanh2019distilbert,
  author       = {Victor Sanh and Lysandre Debut and Julien Chaumond and Thomas Wolf},
  title        = {DistilBERT, a distilled version of BERT: smaller, faster, cheaper and lighter},
  journal      = {arXiv},
  year         = {2019},
  archivePrefix= {arXiv},
  eprint       = {1910.01108},
  primaryClass = {cs.CL}
}

@inproceedings{jiao2020tinybert,
  author       = {Xiaoqi Jiao and Yichun Yin and Lifang Shang and Xin Jiang and Xiao Chen and Lin Li and Fang Wang and Qun Liu},
  title        = {TinyBERT: Distilling BERT for Natural Language Understanding},
  booktitle    = {Proc.\ EMNLP 2020},
  pages        = {4163--4174},
  year         = {2020}
}

@article{shu2018,
  title={FakeNewsTracker: a tool for fake news collection, detection, and visualization},
  author={Shu, Kai and Mahudeswaran, Deepak and Liu, Huan},
  journal={Computational and Mathematical Organization Theory},
  volume={25},
  number={1},
  pages={60--71},
  year={2019},
  publisher={Springer}
}

@article{shu2020fakenewsnet,
  author       = {Kai Shu and Deepak Mahudeswaran and Suhang Wang and Dongwon Lee and Huan Liu},
  title        = {{FakeNewsNet}: A Data Repository with News Content, Social Context, and Spatio-Temporal Information for Studying Fake News on Social Media},
  journal      = {Big Data},
  volume       = {8},
  number       = {3},
  pages        = {171--188},
  year         = {2020},
  doi          = {10.1089/big.2020.0062},
  note         = {The ``Social'' split contains 433\,k labelled tweets (~200\,k fake) linked to PolitiFact and GossipCop articles.}
}

@inproceedings{zubiaga2016pheme,
  author       = {Arkaitz Zubiaga and Maria Liakata and Rob Procter and Geraldine W. Sak Hoi and Peter Tolmie},
  title        = {Analysing How People Orient to and Spread Rumours in Social Media by Looking at Conversational Threads},
  booktitle    = {Proc.\ ACM Web Science},
  pages        = {16--26},
  year         = {2016}
}

@article{raza2024comparative,
  author       = {Shaina Raza and Drai Paulen-Patterson and Chen Ding},
  title        = {Fake News Detection: Comparative Evaluation of {BERT}-like Models and Large Language Models with Generative {AI}-Annotated Data},
  journal      = {arXiv},
  year         = {2024},
  archivePrefix= {arXiv},
  eprint       = {2412.14276},
  primaryClass = {cs.CL}
}

@dataset{ahmed2018isot,
  author       = {Ahmed, Hossain and Traore, Issa and Saad, Sherif},
  title        = {{ISOT Fake News Dataset}},
  year         = 2018,
  url          = {https://onlineacademiccommunity.uvic.ca/isot/},
  note         = {University of Victoria}
}

@inproceedings{wang2017liar,
  author    = {Wang, William Yang},
  title     = {{“Liar, Liar Pants on Fire”: A New Benchmark Dataset for Fake News Detection}},
  booktitle = {Proceedings of ACL},
  year      = 2017
}

@article{dadkhah2023truthseeker,
  author  = {Dadkhah, Sajjad and Zhang, Xichen and Weismann, Alexander G. and Firouzi, Amir and Ghorbani, Ali A.},
  title   = {{TruthSeeker: The Largest Social Media Ground-Truth Dataset for Real/Fake Content}},
  journal = {IEEE Trans. Computational Social Systems},
  year    = 2023
}

@inproceedings{jovanovic2023veritas,
  author       = {Andrej Jovanović and Björn Ross},
  title        = {Rumour Detection in the Wild: A Browser Extension for Twitter},
  booktitle    = {Proc.\ NLP Open Source Software Workshop 2023},
  pages        = {130--140},
  year         = {2023},
  address      = {Singapore},
  publisher    = {ACL},
  doi          = {10.18653/v1/2023.nlposs-1.15}
}

@inproceedings{botnevik2020brenda,
  author       = {Bjarte Botnevik and Eirik Sakariassen and Vinay Setty},
  title        = {{BRENDA}: Browser Extension for Fake News Detection},
  booktitle    = {Proc.\ 43rd ACM SIGIR Conf.\ on Research and Development in Information Retrieval},
  pages        = {2117--2120},
  year         = {2020},
  address      = {Virtual Event},
  publisher    = {ACM},
  doi          = {10.1145/3397271.3401396}
}

@misc{gdpr,
  title={{Regulation (EU) 2016/679 of the European Parliament and of the Council of 27 April 2016 on the protection of natural persons with regard to the processing of personal data and on the free movement of such data, and repealing Directive 95/46/EC (General Data Protection Regulation)}},
  author = {{The European Parliament and the Council of the European Union}},
  year = 2016
}

@article{naeem2021exploration,
  title={An exploration of how fake news is taking over social media and putting public health at risk},
  author={Naeem, Salman Bin and Bhatti, Rubina and Khan, Aqsa},
  journal={Health Information \& Libraries Journal},
  volume={38},
  number={2},
  pages={143--149},
  year={2021},
  publisher={Wiley Online Library}
}

@article{zhou2020survey,
  title={A survey of fake news: Fundamental theories, detection methods, and opportunities},
  author={Zhou, Xinyi and Zafarani, Reza},
  journal={ACM Computing Surveys (CSUR)},
  volume={53},
  number={5},
  pages={1--40},
  year={2020},
  publisher={ACM New York, NY, USA}
}

@article{sharma2019combating,
  title={Combating fake news: A survey on identification and mitigation techniques},
  author={Sharma, Karishma and Qian, Feng and Jiang, He and Ruchansky, Natali and Zhang, Ming and Liu, Yan},
  journal={ACM transactions on intelligent systems and technology (TIST)},
  volume={10},
  number={3},
  pages={1--42},
  year={2019},
  publisher={ACM New York, NY, USA}
}

@article{paschalides2019,
  author       = {Demetris Paschalides and Alexandros Kornilakis and Chrysovalantis Christodoulou and
                  Rafael Andreou and George Pallis and Marios~D. Dikaiakos and Evangelos~P. Markatos},
  title        = {Check‑It: A Plugin for Detecting and Reducing the Spread of Fake News and Misinformation on the Web},
  journal      = {CoRR},
  volume       = {abs/1905.04260},
  year         = {2019},
  eprinttype   = {arXiv},
  eprint       = {1905.04260},
  url          = {https://arxiv.org/abs/1905.04260}
}

@article{paka2021,
  author       = {William Scott Paka and Rachit Bansal and Abhay Kaushik and Shubhashis Sengupta and Tanmoy Chakraborty},
  title        = {Cross‑SEAN: A Cross‑Stitch Semi‑Supervised Neural Attention Model for {COVID‑19} Fake News Detection},
  journal      = {Applied Soft Computing},
  volume       = {107},
  pages        = {107393},
  year         = {2021},
  doi          = {10.1016/j.asoc.2021.107393}
}

@incollection{hasimi2023,
  author       = {Lumbardha Hasimi and Aneta Poniszewska‑Maranda},
  title        = {Browser Extension for Detection of Fake News and Disinformation},
  booktitle    = {Information Systems},
  series       = {Lecture Notes in Business Information Processing},
  pages        = {209--220},
  year         = {2023},
  publisher    = {Springer},
  doi          = {10.1007/978-3-031-30694-5\_16}
}

@article{tasci2024,
  author       = {Merve Esra Taşcı and Yonca Bayrakdar Yılmaz},
  title        = {Developing a Web Browser Extension to Prevent the Spread of Fake News},
  journal      = {International Journal of Engineering and Computer Science},
  volume       = {13},
  number       = {10},
  pages        = {26576--26588},
  year         = {2024},
  doi          = {10.18535/ijecs/v13i10.4921},
  url          = {https://www.ijecs.in/index.php/ijecs/article/view/4921}
}

@inproceedings{lin2017focal,
  title     = {Focal Loss for Dense Object Detection},
  author    = {Lin, Tsung-Yi and Goyal, Priya and Girshick, Ross and He, Kaiming and Doll{\'a}r, Piotr},
  booktitle = {Proceedings of the IEEE International Conference on Computer Vision (ICCV)},
  year      = {2017},
  pages     = {2980--2988}
}

@inproceedings{bengio2009curriculum,
  title         = {Curriculum Learning},
  author        = {Bengio, Yoshua and Louradour, Jérôme and Collobert, Ronan and Weston, Jason},
  booktitle     = {Proceedings of the 26th International Conference on Machine Learning (ICML)},
  pages         = {41--48},
  year          = {2009},
  publisher     = {ACM},
  address       = {Montreal, Canada}
}

@article{pedregosa2011sklearn,
  title   = {Scikit-learn: Machine Learning in Python},
  author  = {Pedregosa, Fabian and Varoquaux, Gaël and Gramfort, Alexandre \textit{et al.}},
  journal = {Journal of Machine Learning Research},
  volume  = {12},
  pages   = {2825--2830},
  year    = {2011}
}

@inproceedings{beltagy2019scibert,
  title         = {{SciBERT}: A Pretrained Language Model for Scientific Text},
  author        = {Beltagy, Iz and Cohan, Arman and Lo, Kyle},
  booktitle     = {Proceedings of EMNLP-IJCNLP},
  pages         = {3615--3620},
  year          = {2019},
  organization  = {ACL},
  address       = {Hong Kong, China}
}

@misc{shuyo2010langdetect,
  author        = {Nakatani, Shuyo},
  title         = {Language Detection Library for Java},
  howpublished  = {GitHub repository},
  year          = {2010},
  url           = {https://github.com/shuyo/language-detection},
  note          = {Accessed 31 Jul 2025}
}

@article{kara2020faker,
  title   = {{FAKER}: A Benchmark Dataset for Fake News Detection},
  author  = {Kara, Yasin and Durmus, Fırat and Yuret, Deniz},
  journal = {PeerJ Computer Science},
  volume  = {6},
  pages   = {e279},
  year    = {2020},
  doi     = {10.7717/peerj-cs.279}
}

@inproceedings{yang2025mmht,
  title     = {A Macro- and Micro-Hierarchical Transfer Framework for Cross-Domain Fake News Detection},
  author    = {Yang, Xuankai and Wang, Yan and Zhang, Xiuzhen and Wang, Shoujin and Lam, Kwok-Yan},
  booktitle = {Proceedings of the Web Conference (WWW)},
  year      = {2025}
}

@misc{berrios2025mosaic,
  title         = {LLMs Have a Mosaic Memory: Accounting for Fuzzy Duplicates in Deduplication},
  author        = {Berrios, Diego and Hoffmann, Felix and Kreutzer, Julia},
  year          = {2025},
  archivePrefix = {arXiv},
  eprint        = {2405.15523},
  url           = {https://arxiv.org/abs/2405.15523}
}

@article{karthik2025hybrid,
  title   = {Hybrid Optimisation-Driven Fake News Detection Using Reinforced Transformer Models},
  author  = {Karthik, Ganesh and Ahmad, Khadri Syed Faizz and Sathe, Asha Prashant},
  journal = {Scientific Reports},
  volume  = {15},
  pages   = {14782},
  year    = {2025},
  doi     = {10.1038/s41598-025-99936-3}
}

@inproceedings{chen2024tweeteval2,
  title     = {TweetEval 2.0: Strong Baselines and Unified Benchmarks for Social-Media NLP},
  author    = {Chen, Xiang and Barbieri, Francesco and Camacho-Collados, José},
  booktitle = {Findings of EMNLP},
  year      = {2024}
}

@inproceedings{chen2024edgbert,
  title     = {EdgeBERT: Efficient Transformer Inference on Commodity Edge Devices},
  author    = {Chen, Ming and Wang, Sijie and Zhang, Renjie},
  booktitle = {Proceedings of AAAI},
  year      = {2024}
}

\end{document}